\documentclass[aps,prc,superscriptaddress,twocolumn]{revtex4-2}
\usepackage{color}
\usepackage{graphicx}

\usepackage[dvipsnames]{xcolor}

\usepackage{cancel}

\newcommand{\beq}{\begin{equation}}
\newcommand{\eeq}{\end{equation}}

\usepackage[reqno]{amsmath}
\usepackage{amsthm}
\usepackage{amssymb}

\begin{document}

\title{Side Ridge
in Ar + KCl Collisions at $1.8 \, \text{GeV/nucleon}$ with Reaction-Plane Deblurring}

\author{Pawel Danielewicz}
\email[]{daniel@frib.msu.edu}
\affiliation{Facility for Rare Isotope Beams and
Department of Physics and Astronomy, Michigan State University, East Lansing, Michigan 48824, USA}

\author{Herbert Str\"{o}bele}
\affiliation{Institut f\"{u}r Kernphysik, Goethe-Universit\"{a}t, 60438 Frankfurt, Germany}

\author{Pierre Nzabahimana}
\affiliation{Facility for Rare Isotope Beams and
Department of Physics and Astronomy, Michigan State University, East Lansing, Michigan 48824, USA}

\date{\today}

\begin{abstract}
Reaction-plane deblurring is applied to the triple-differential distributions of protons, deuterons and negatively charged pions from central Ar + KCl collisions at $1.8 \, \text{GeV/nucleon}$, measured with the LBL-GSI Streamer Chamber, to yield distributions relative to the true reaction plane of the collisions. Within the reaction plane and in the forward cm rapidity region a side ridge away from the beam axis is observed: the distributions peak away from the beam direction for protons and deuterons but not for pions. These findings are consistent with a source of particles moving at an in-plane transverse velocity of $\sim 0.1 \, c$.  Transport pBUU calculations yield results in semi-quantitative agreement with those from the data.
\end{abstract}

\maketitle
In Heavy-Ion Collisions (HIC) \cite{danielewicz_determination_2002} and in neutron star mergers \cite{Ligo2017} nuclear matter gets compressed to densities higher than normal for interiors of nuclei. In both classes of the events, respectively highly microscopic and macroscopic, the response to matter compression drives the system dynamics.  In observing the events one hopes to learn, in particular, about the nuclear Equation of State (EoS)~\cite{the_ligo_scientific_collaboration_and_the_virgo_collaboration_gw170817:_2018, huth_constraining_2022, lynch_decoding_2022}.  For the microscopic systems at intermediate collision energies, where few new particles are produced, the use of hydrodynamics has been though on the extreme side, in employing EoS to close the set of local conservation laws of baryon number, momentum and energy.  The reason is in ignoring or approximating, in terms of transport coefficients, the effect of the finite length of nuclear mean free path that is significant relative to the system size.  Yet early on, hydrodynamics has been employed for intermediate-energy HIC, on account of its simplicity~\cite{stocker_jets_1982, stocker_high_1986}.  More realistic transport descriptions have been developed for HIC over time, with nucleons moving under the influence of a mean field due to other nucleons, engaging in elementary collisions and producing mesons and baryon resonances \cite{bertsch_guide_1988, wolter_transport_2022}.

In HIC modelling, hydrodynamics produced dramatic collective effects and, in particular, deflection of the so-called participant matter, emerging from the compression zone, in the reaction plane away from the beam axis~\cite{stocker_high_1986}.   In the triple-differential distribution associated with the reaction plane, that conventionally becomes the $xz$ plane, the emission probability for particles moving forward in the center of mass (cm) (forward is relative to beam direction in fixed target experiments) maximized away from the beam axis, even in the light system  $^{40}$Ar + $^{40}$Ca at $393 \, \text{MeV/nucleon}$~\cite{stocker_jets_1982}.  Observation of such a side ridge was claimed by the Plastic Ball Collaboration for central $400 \, \text{MeV/nucleon}$ $^{93}$Nb + $^{93}$Nb collisions~\cite{gustafsson_collective_1984}.  The~authors named this phenomenon a bounce-off in tying it to projectile remnants.  The collision events in~\cite{gustafsson_collective_1984} were rotated to a common estimated reaction plane and the resulting triple-differential particle distribution was examined in that plane.  What was not recognized at the time was that the use of a particle in determining the reaction plane and then relating the particle to that plane could produce fake signals which could be attributed to the reaction-plane effects.  As the issue was recognized and corrected for \cite{danielewicz_transverse_1985}, the attention was turned to azimuthal moments associated with the reaction plane \cite{voloshin_flow_1996, poskanzer_methods_1998}.  For those moments to be finite, including the one being behind finite flow angles, it is sufficient for azimuthal distributions around the beam axis to be anisotropic.  On their own, they do not imperil the perception of overall persistence of motion along the beam direction in the final states of energetic HIC.  That persistence may be expected given the energies of collisions far in excess of other pertinent energy scales in nuclei and the mean free paths that cannot be dismissed relative to system size.

In recent years, attention came back to the triple-differential distributions associated with the reaction plane.  Relying on high-event statistics and associated ability to find high angular moments of azimuthal distributions relative to the reaction plane, the HADES Collaboration attempted to assess the shape of proton emission relative to the reaction plane in $1.23 \, \text{GeV/nucleon}$ Au + Au collisions \cite{hades_collaboration_directed_2020}.  Working on behalf of the S$\pi$RIT Collaboration, Danielewicz and Kurata-Nishimura proposed to employ reaction-plane deblurring to arrive at triple-differential distributions associated with the reaction plane \cite{danielewicz_deblurring_2022}.   Here, we employ that last strategy to  demonstrate the existence of a side ridge in central ($b \lesssim 2.4 \, \text{fm}$) Ar + KCl collisions at $1.8 \, \text{GeV/nucleon}$, measured by the LBL Streamer Chamber Collaboration almost 40 years ago~\cite{strobele_charged-particle_1983}.  These data, containing just 495 events, originally turned the attention in HIC to azimuthal moments relative to the reaction plane~\cite{danielewicz_transverse_1985}.  If a side-ridge can be identified in such a light system at a~relatively high energy, with such a low event statistics, it is perceivable that it is possible to see it under broad circumstances in intermediate-energy HIC.  In the following, we discuss in sequence the reaction-plane deblurring, analyze data, compare the results to transport-model calculations and conclude.

In the reaction-plane deblurring, triple-differential distributions are first established relative to the estimated azimuthal direction~\cite{danielewicz_deblurring_2022} of the reaction-plane around the beam axis.  Different sensible constructs may be used for estimating the plane direction, but the particle or particles that get refereed to the plane must be excluded  from the construct, to avoid the self-correlation.  It is in particular common to estimate the direction of the reaction plane with a transverse~$\mathbf{Q}$ vector that is the sum of weighted transverse momenta $\mathbf{p}^\perp$ of measured particles~\cite{danielewicz_transverse_1985}.  With the above, for particle $\mu$, the reaction plane direction would be then estimated with
\begin{equation}\label{eq:Qmu}
\mathbf{Q}_\mu = \sum_{\nu \ne \mu} \omega_\nu \, \mathbf{p}_\nu^\perp \, .
\end{equation}
Here, $\omega_\nu$ are weights taken as positive for particle types and rapidity $y$ and transverse momentum $p^\perp$ regions where the vector momenta correlate with the positive direction of the reaction plane, negative when they correlate with the negative direction and zero if no pronounced correlation is expected.  In a symmetric or near-symmetric system, the weights may be chosen for baryons as
\begin{equation}\label{eq:wnu}
  \omega_\nu = \begin{cases}
                 +1, & \mbox{if } y_\nu^* > \delta \, , \\
                 -1, & \mbox{if } y_\nu^* < -\delta \, , \\
                 0, & \mbox{otherwise}.
               \end{cases}
\end{equation}
Here, $y^*$ is the rapidity in the nucleon-nucleon (NN) center of mass (cm) and $\delta$ is a cut-off eliminating contributions from  midrapidity particles for which transverse momenta neither well correlate nor anticorrelate with the reaction plane.  For pions the weights may be taken as~0 irrespectively of the rapidity, as these paricles, exhibiting weak space-momentum correlations, are expected to also exhibit weak correlations with the reaction plane.

\begin{figure}
\centering
\vspace*{1.5ex}
\includegraphics[width=.75\linewidth]{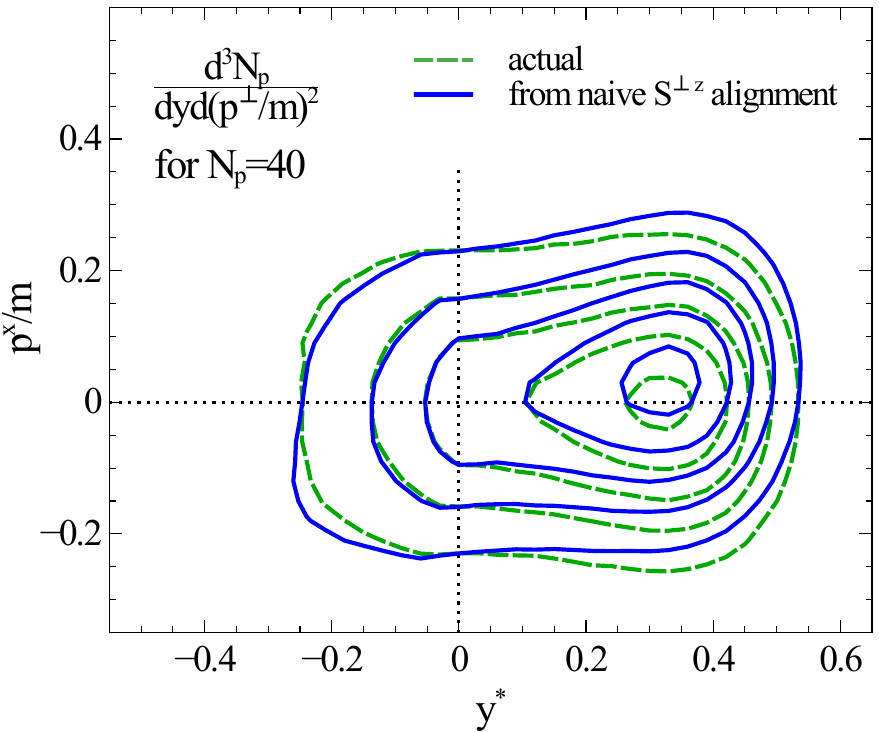}
\vspace*{-2.5ex}
\caption{ Contour plot of 
the yield of protons in the reaction plane in a schematic simulation with $N_p=40$ protons generated with a similar phase space distribution as recorded in the $400 \, \text{MeV/nucleon}$ $^{93}$Nb + $^{93}$Nb experiment~\cite{gustafsson_collective_1984}.  The dashed contours represent the input distribution lacking any correlation with the reaction plane.  The solid contours represent the distribution obtained from aligning reaction planes estimated with the flow tensor constructed from all protons.  The contours represent values of the distribution in multiples of 100, increasing from outside in. The dotted lines serve as a reference.}
\label{fig:dnp}
\end{figure}

To illustrate what happens when a particle is referred to the reaction plane estimated with a construct where the particular particle was employed ($\nu \ne \mu$ condition dropped from Eq.~\eqref{eq:Qmu}), we carry out a schematic simulation of emission with particles lacking any correlation with each other, including any through the reaction plane.  Specifically, we take the characteristics of emission to be similar to those in $400 \, \text{MeV/nucleon}$ $^{93}$Nb + $^{93}$Nb collisions, as studied by the Plastic Ball Collaboration~\cite{gustafsson_collective_1984}.  However, for simplicity, we take $N_p=40$ protons only, emitted independently in each event in an axially symmetric manner around the beam axis, with a suppression of detection around the target.  The distribution of the number of generated protons as function of in-plane transverse momentum and rapidity is represented with dashed contours in Fig.~\ref{fig:dnp}. (For the upper half of the plot only a small slice around $\phi=0$ and for the lower half around $\phi=\pm 180^\circ$ was used.)  At any rapidity the distribution maximizes along the beam direction, i.e., the $p^\perp = 0$ line.  When the reaction plane is estimated with the cm flow tensor $S^{ij}= \sum_\nu {p^i_\nu \, p^j_\nu}/{2m_\nu}$, equivalent to the use of $\omega_\nu = p^z_\nu/2m_\nu$ in \eqref{eq:Qmu}, and the events are rotated to have their estimated reaction planes aligned, the resulting distribution of the protons in the estimated reaction plane becomes the one indicated with solid contours in Fig.~\ref{fig:dnp}.   At any nonzero rapidity, $y^* \ne 0$, the triple-differential distribution maximizes at a finite $p^x$, positive at $y^*>0$ and negative at $y^*<0$, with a pile-up away from the beam axis, even though there is none in the simulation ($p^x$ is the transverse momentum component in the direction of the reaction plane). The~position of the maximum at forward rapidities is similar to that in the Plastic Ball observation~\cite{gustafsson_collective_1984}. This is not to say that the side ridge (bounce-off for projectile remnants) is absent from the Plastic Ball data, only that the procedure applied there was not optimal to filter the phenomenon out.  The flow observable which uses the peaking of the flow tensor distribution at a finite angle away from the beam axis, as in Fig.~1 of~\cite{gustafsson_collective_1984} is not affected by such a bias. For the flow-angle peak, it suffices to have an anisotropy of the single particle distributions tied to the reaction plane, as for example evidenced in a finite $\langle p^x(y) \rangle$~\cite{danielewicz_transverse_1985, doss_nuclear_1986}. A pile-up away from the beam axis is not a necessary condition.

\begin{figure*}
  \centering
\hspace*{.005\linewidth}\includegraphics[width=.31\linewidth]{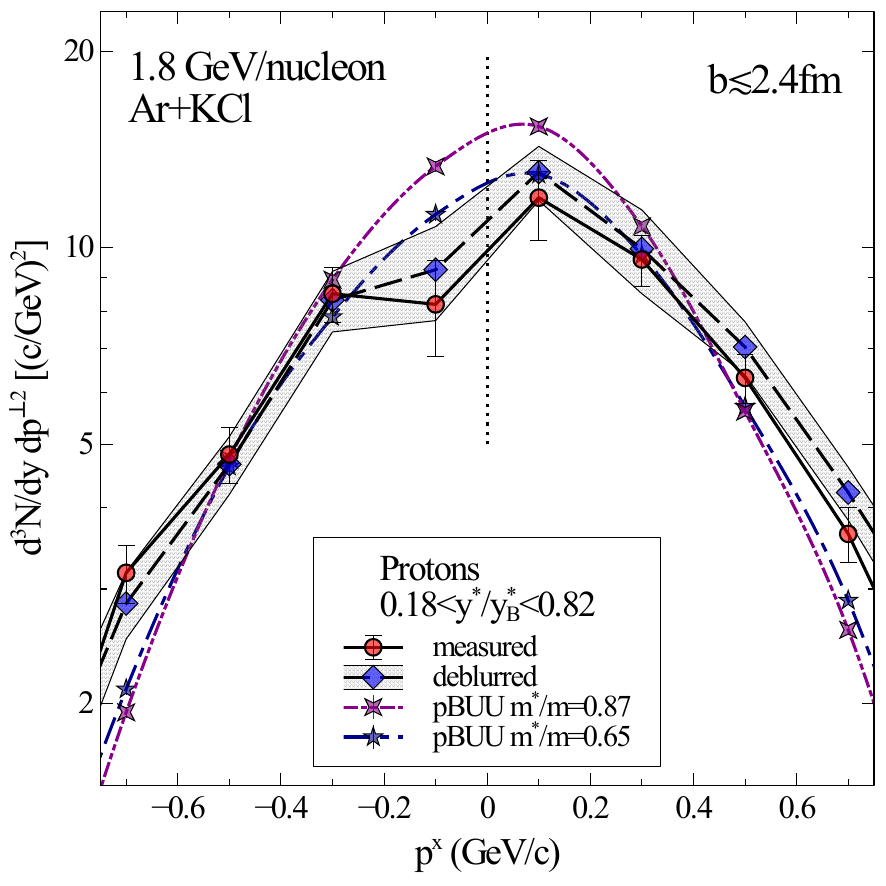}\hspace*{.01\linewidth}
\includegraphics[width=.31\linewidth]{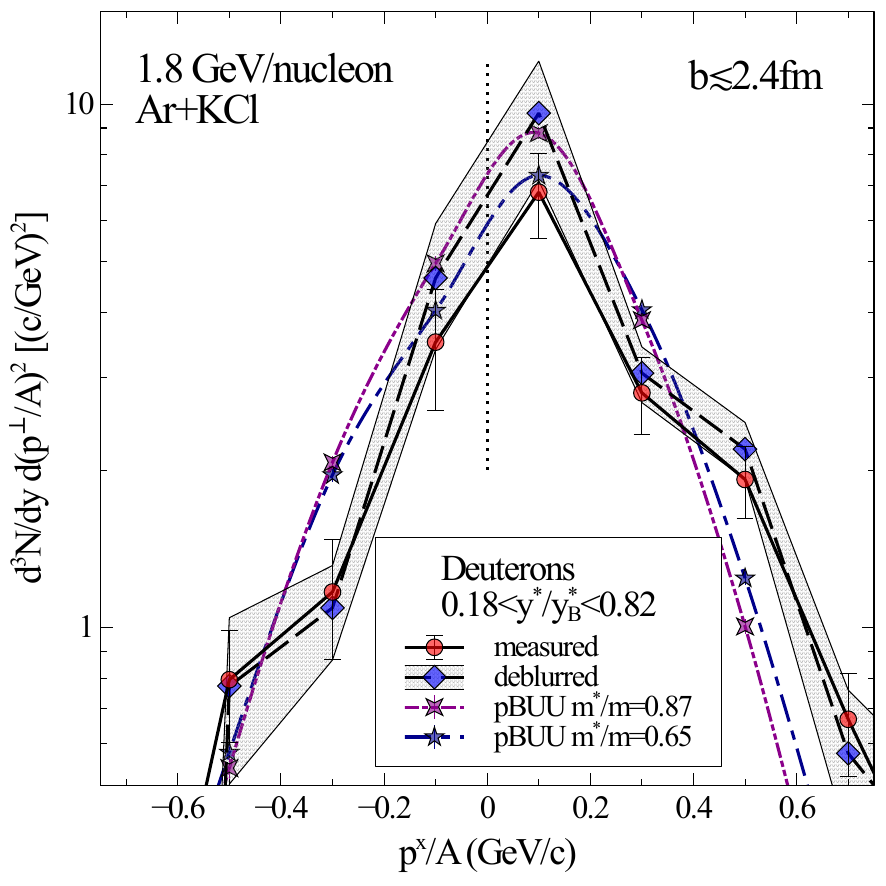} \hspace*{.01\linewidth}  \includegraphics[width=.31\linewidth]{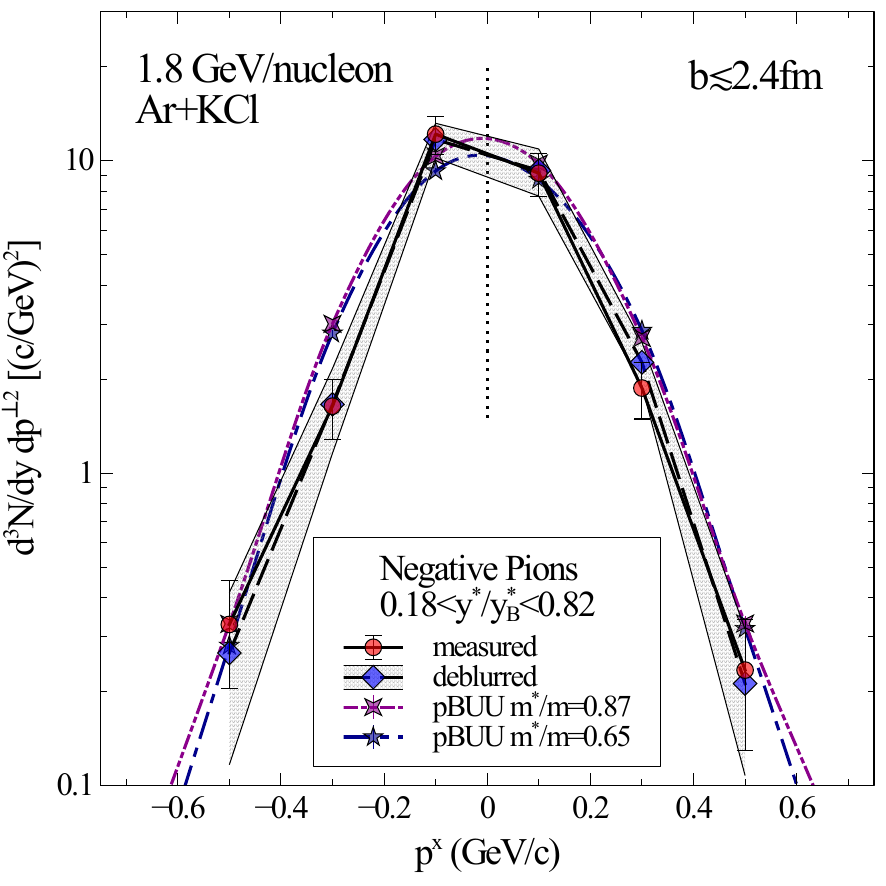}
  \hspace*{.025\linewidth}
  \caption{
  Invariant 
  yields of protons, deuterons and negative pions, from left to right, as a function of the momentum component in the reaction plane, within the forward rapidity region, $0.18 < y^*/y_B^* < 0.82$, for the $1.8 \, \text{GeV/nucleon}$ Ar + KCl reaction~\cite{strobele_charged-particle_1983, danielewicz_transverse_1985} at impact parameters $b \lesssim 2.4 \, \text{fm}$. (Only the yields in the azimuthal bins surrounding $\phi=0$ and $\phi=\pm 180^\circ$ are considered.)  The circles represent experimental results from aligning estimated reaction planes for individual particles.  The diamonds represent results from deblurring the impact of the spread of the estimated reaction plane around the true plane, on the measured distributions.  The squashbox and star symbols represent pBUU~\cite{danielewicz_determination_2000} results with, respectively, a weaker and stronger momentum dependence in the baryon mean fields.  Lines guide the eye. 
   }  \label{fig:3Dif}
\end{figure*}

In analyzing the $1.8 \, \text{GeV/nucleon}$ Ar + KCl data~\cite{strobele_charged-particle_1983, danielewicz_transverse_1985}, we rotate individual particles from all events to align the reaction planes estimated for each particle separately. 
The LBL-GSI Streamer Chamber provides good identification of protons, deuterons and negative pions in the forward cm hemisphere of this reaction \cite{strobele_kernmaterie_1986}.  With low events statistics of just 495 and multiplicities limited by the system size, we use a coarse binning in the azimuthal angle relative to the reaction plane, dividing the $360^\circ$ range into 16 intervals.  In transverse momentum magnitude, we take $200 \, \text{MeV}/c$ intervals for the pions and the same span, per baryon number, for the baryons.  In rapidity, we take a relatively wide range centered at half of the beam rapidity $y_B^*$ in the NN cm: $0.18 < y^*/y_B^* < 0.82$.   With the latter, we exclude both the mid-rapidity particles, that by symmetry are not likely to contribute to a side ridge, and any potential spectator contributions.  The invariant yields of protons, deuterons and negative pions, as a function of momentum component in the plane of the estimated reaction plane are shown with circles in Fig.~\ref{fig:3Dif}.
A significant asymmetry, with a pile-up away from the beam axis, may be seen in the distributions represented with circles, between the positive (direction of $\bf Q$) and negative sides of $p^x$, for baryons, but not for pions. The level of asymmetry can be expected to depend on the reaction-plane resolution. Specifically, the distribution measured relative to the estimated reaction plane can be written in terms of the distribution relative to the true reaction plane as~\cite{danielewicz_deblurring_2022}
\begin{equation}\label{eq:dN}
  \frac{d^3N}{p^\perp d p^\perp \, dy \, d\phi'} = \int d\phi \, \frac{dP}{d\phi'}\big(\phi'| \phi   \big) \, \frac{d^3N}{p^\perp d p^\perp \, dy \, d\phi} \, .
\end{equation}
Here, the l.h.s.\ particle distribution is in the angle~$\phi'$ relative to the estimated plane and the r.h.s.\ distribution is in the angle~$\phi$ with respect to the true plane.  The conditional probability density $dP/d\phi'$ on the r.h.s.\ is for a particle that has an angle~$\phi$ getting detected at an angle~$\phi'$.  The~narrower $dP/d\phi'$, the more pronounced will be any azimuthal features in $dN/d\phi'$ and, if $dP/d\phi'$ were to be isotropic, then $dN/d\phi'$ would become isotropic too.  Eq.~\eqref{eq:dN} is a blurring relation, with $dP/d\phi'$ acting as the blurring function.  The blurring relation may include effects of detector efficiency, but, with the experimental statistics at hand, directly relevant systematic errors are difficult to assess and corrections for those errors are expected to be small compared to the errors of statistical origin for the triple-differential data we have obtained.

When the fluctuation of the vector $\mathbf{Q}$ around its average $\langle Q^x \rangle \, \hat{\mathbf{e}}^x$ directed in the reaction plane is the only factor in the blurring, then the blurring function becomes dependent only on the difference of the azimuthal angles, $\Delta \phi = \phi'-\phi$.  When multiplicity is large and central limit can be invoked, yielding Gaussian fluctuations of $\mathbf{Q}$ around its average, the blurring function can be given in an analytic form \cite{voloshin_flow_1996, danielewicz_deblurring_2022}:
\begin{equation}\label{eq:dP}
\frac{\text{d}P}{\text{d}\phi'}  \simeq \alpha  \big\lbrace 1 + \sqrt{\pi} \beta  \exp{(\beta^2)}
 \big[ 1 + \text{erf} ( \beta  )  \big] \big\rbrace \, .
\end{equation}
Here, $ \alpha$ is a normalization constant, $  \beta = \frac{\langle Q^x \rangle}{\sigma_\mathbf{Q}}  \cos \Delta \phi $, and $\sigma_\mathbf{Q}^2 = \langle (\mathbf{Q} - \langle Q^x \rangle \, \hat{\mathbf{e}}^x )^2 \rangle$.  In our simulations the central limit works well for the blurring function at multiplicities of particles contributing to $\mathbf{Q}$ as low as 10.  The averages needed in \eqref{eq:dP} can be estimated using averages of scalars from particle momenta, squares and scalar products~\cite{danielewicz_collective_1988, voloshin_flow_1996}. Eq.~\eqref{eq:dP} is commonly employed in finding renormalization factors for the Fourier coefficients of azimuthal distributions~\cite{voloshin_flow_1996}, turning $v_n' = \langle \cos {n\phi'} \rangle$ into $v_n = \langle \cos {n\phi} \rangle$.

Notably, the Gaussian of the distribution of $\mathbf{Q}$ in the plane perpendicular to the beam axis was  tacitly assumed to be isotropic for Eq.~\eqref{eq:dP}.  A possible anisotropy can again be assessed with averages of scalars from momenta~\cite{danielewicz_deblurring_2022}, and none could be detected.  For a light system at high energy a negligible anisotropy is expected, and transport theory agrees with that for the particular system, so we put that anisotropy to zero, as standard in the literature \cite{voloshin_flow_1996}.  Otherwise, for our sample we find $\frac{\langle Q^x \rangle}{ \sigma_\mathbf{Q}} \simeq \frac{2.33\, \text{GeV}/c}{2.93 \, \text{GeV}/c} = 0.80$ needed for Eq.~\eqref{eq:dP}.

To gain information about the triple-differential distribution associated with the true reaction plane, we work to remove the effect of blurring from the measured distribution directly, or apply deblurring to the measured distribution, rather than work with individual Fourier coefficients.  Still the deblurring can be understood in terms of an expansion too, but that in singular value decomposition for the discretized blurring function in \eqref{eq:dN}.  The~deblurring relies on the dominance of singular vectors connected to the largest singular values in physical distributions~\cite{hansen_deblurring_2006, mamba23}, and these vectors, in turn, are tied to the slowest variation in the azimuthal angle.  Contributions of these vectors are restored faithfully.  A regularization is applied, e.g., \cite{remmele_vector_2009, dey_richardsonlucy_2006}, to prevent an uncontrolled rise of contributions from vectors connected to very small singular values, that are tied to the fastest variations and are expected to make small contributions to physical results.  The rise of such contributions could be triggered by noise in measurements or numerical fluctuations during the deblurring.

The deblurring method we actually apply is the iterative Richardson-Lucy method~\cite{richardson_bayesian-based_1972, lucy_iterative_1974}, which is based on Bayes' theorem and has a long record in optical imaging~\cite{vankawala_survey_2015}. The method exploits the discretization that we already introduced, and we apply the regularization of Refs.~\cite{remmele_vector_2009, dey_richardsonlucy_2006, danielewicz_deblurring_2022} with $\lambda=2.4 \%$.  The regularization, that can result in weak correlations between the bins~\cite{nzabahimana_deconvoluting_2023}, is applied both in the direction of azimuthal angle and transverse momentum, as noise differentiating $p^\perp$ bins could also get amplified in the deblurring.  To estimate errors in the deduced distribution relative to the true plane, we resample the measured distribution with its errors and repeat the deblurring~\cite{nzabahimana_deconvoluting_2023}.

Fig.~\ref{fig:3Dif} shows the yields of protons, deuterons, and negatively charged pions  in the reaction plane ($\phi = 0^\circ/\pm 180^\circ$), as a function of the transverse momentum component $p^x$ before (circles) and after (diamonds) deblurring.  On average the deblurring enhances the present anisotropies relative to the reaction plane, moving the yields up or down.  For baryons, the $p^x>0$ side is generally enhanced  relative to $p^x<0$ already for the measured distributions.  On average the deblurring lifts the $p^x>0$ side and depletes $p^x<0$.  For pions, with very little if any anisotropy evident relative to the estimated reaction plane, no significant change is produced by the deblurring.  The part of the side ridge of the baryon distributions which peaks away from the beam in the selected rapidity rapidity interval at $p^x/A \sim 100~\text{MeV}/c$, suggest a source in the forward cm hemisphere moving at a transverse collective velocity of $0.1 \, c$.  The lack of a shift of the maximum in the pion spectrum in Fig.~\ref{fig:3Dif} (right-side panel) supports the interpretation of the shift in the baryon spectra in terms a collective velocity of the source of matter, since the pions, generally moving relativistically with respect to the surrounding matter, exhibit a weak sensitivity to any collective velocity field.

\begin{figure}
\centering
\vspace*{1.5ex}
\includegraphics[width=.66\linewidth]{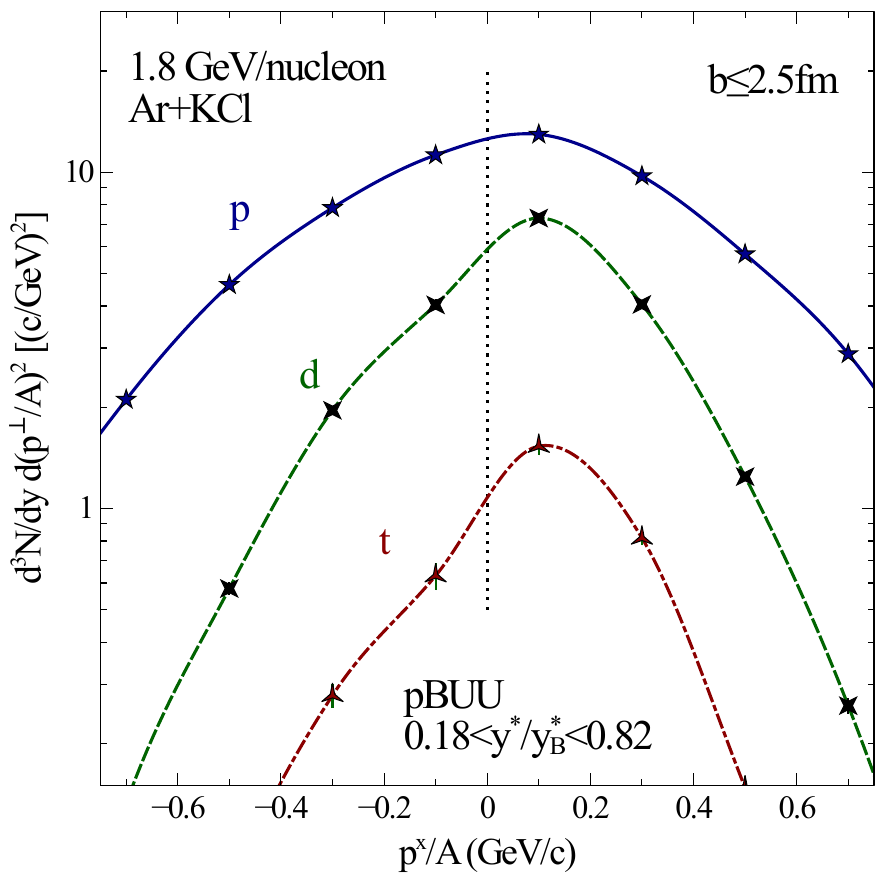}
\vspace*{-2.5ex}
\caption{Yields of protons (5-arm stars), deuterons (squashboxes) and tritons (3-arm stars), from pBUU calculations~\cite{danielewicz_determination_2000} with $m^*/m=0.65$, as a function of the momentum component in the reaction plane, within the forward rapidity region, $0.18 < y^*/y_B^* < 0.82$, for the $1.8 \, \text{GeV/nucleon}$ Ar + KCl reaction at impact parameters $b \le 2.5 \, \text{fm}$.  Lines guide the eye.}
\label{fig:H}
\end{figure}

Evidence for the pile-up away from the beam axis in a light system at a relatively high incident energy may be surprising and we thus supplement our results from measurements with those from the transport simulations in the pBUU model~\cite{danielewicz_determination_2000}.  The model yields results that are close to those from the data.  The pBUU are represented in Fig.~\ref{fig:3Dif} with a squashbox and star, respectively, for a weaker ($m^*/m=0.87$ at the Fermi momentum in normal matter) and stronger ($m^*/m = 0.65$ at the same) momentum dependence in nucleonic mean field.  The data seem to favor, marginally, the stronger momentum dependence.  Notably, for such a light system we find no significant sensitivity to the exclusive variation of density dependence for the mean field, within typical range of testing within HIC modelling.  This seems to parallel the situation with heavier systems at semi-peripheral impact parameters, where sensitivity to the density dependence is lost, but that to the momentum dependence is kept \cite{danielewicz_determination_2000}.  We complement the results here showing the predictions of pBUU for all three hydrogen isotopes in Fig.~\ref{fig:H}.  The predicted triton spectra also peak at $ p^x/A \sim 100 \, \text{MeV}/c $, consistently with the interpretation of an $x$-component of collective velocity at the value of about $0.1 \, c$.

To sum up, we have examined triple-differential particle distributions from Ar + KCl collisions at $1.8 \, \text{GeV/nucleon}$, measured in the LBL-GSI Streamer Chamber.  The deblurring applied to the distributions removes the impact of uncertainty in the reaction-plane determination dependent on a particular detection system, giving the triple-differential distributions a status similar to the Fourier coefficients $v_n$ used for collision data in the past.  The distributions reveal the presence of a side ridge characterized by a transverse collective velocity of~$0.1 \, c$ at large rapidity.  The pBUU predictions agree with the inferences from data at a semi-quantitative level.  Otherwise, model calculations, here and prior to~\cite{danielewicz_deblurring_2022, reichert_3-dimensional_2022}, yield nontrivial features in the predicted triple-differential yields beyond a side ridge, such as differences in the in-plane participant and spectator shapes of the distributions.  These could be searched for in higher statistics data, in addition to the obvious expectation of finding an improved EoS resolution for heavy collisions systems.

\begin{acknowledgements}
The authors thank Grazyna Odyniec for useful comments. This work was supported by the U.S.\ Department of Energy Office of Science under Grant {DE}-{SC}0019209.
\end{acknowledgements}

\bibliography{splash}

\end{document}